\documentstyle[preprint,aps]{revtex}
\begin{document}
\draft
\preprint{MAD/TH/94-2}
\title
{New solution for dynamical symmetry breaking with\\ top
and bottom quark condensates}
\author{Bernice Durand and Tu Zhang}
\address
{Department of Physics, 1150 University Avenue,\\
University of Wisconsin-Madison, Madison, Wisconsin, 53706}
\maketitle
\begin{abstract}
Starting from a general $\rm SU_2\times U_1$ invariant interaction
Lagrangian ${\cal L}_{\rm int}$ with four-fermion interactions between
top $t$ and bottom $b$
quarks and working in the bubble approximation, we get a single Higgs
as a bound state of quark pairs by allowing both $\langle \bar tt\rangle$
and $\langle\bar bb\rangle$ to be nonzero and the $t$ and $b$ states to mix.
We find relations between the three four-fermion
couplings $g_t, g_b,$ and $g_{tb}$ and show the new result
that they may all be finite, with $g\Lambda ^2\gg1$,
where $\Lambda$ is the cutoff.
Thus the dimensionless couplings $g' = g\Lambda ^2$
correspond to strong interactions.
Previous work with either one or more quark condensates found a fine-tuning
condition giving $g\Lambda ^2\sim$ O($1$).  The Higgs mass $m_H$
is approximately the single-vev value $m_H\approx 2m_t$, and the
quark mass ratio is
$m_t/m_b\approx g_b/g_{tb}$.   There is a new
symmetry of ${\cal L}_{\rm int}$, corresponding to a flat
direction in the space of composite states.
Breaking this symmetry
by turning on one small eigenvalue of the coupling matrix turns on
the quark masses and
introduces the massive Higgs state, so that
$m_b,m_t,m_H\ll \Lambda$ is natural.
This symmetry takes over the role of ``fine tuning" in the
single-quark model or multi-quark two-doublet Higgs model.
A possible regime of interest for this
solution is $m_t \ll \Lambda \sim$ 2 TeV.
\end{abstract}
%12.60.Rc Composite models; 14.80.Cp non standard model Higgs bosons
\section{INTRODUCTION}

The Higgs boson is introduced in the standard model
to implement the breaking of the electroweak symmetry, where it leads
to problems with the high-energy behavior.  Carter and Pagels \cite{cp}
suggested using the Nambu--Jona-Lasinio mechanism \cite{njl} to implement
dynamical symmetry breaking.  Bardeen, Hill, and Lindner \cite{bhl}
investigated using this mechanism by
treating the Higgs as a $t\bar t$ bound state, since the
top quark is so much more massive than the other five quarks, hence closer
to the massive Higgs.  Many authors have carried this idea further
\cite{s}-\cite{lm}.

The $b$ quark, as the $SU(2)_L$ partner of the $t$, can be included in the
dynamical symmetry breaking in a natural way.  This was pointed out by
Bardeen, Hill and Lindner, and done by Suzuki \cite{s}, who carried his
formalism to six quarks.
In the standard model, the coupling of a quark to the Higgs boson
is proportional to the quark's mass, which results in the strongest
coupling for the $t$ and $b$ quarks.  However the inclusion of the
lighter quark pairs is straightforward and makes a nice general
model.

In both \cite{bhl} and \cite{s}, there is a fine-tuning condition needed to
obtain solutions to the mass-gap equations for the bound state(s).
This condition requires that all the effective four-fermion interactions
be tuned to be O($1/\Lambda ^2$), where $\Lambda$ is the cutoff in
the model,
so that $g\Lambda ^2\sim {\rm O}(1)$.

In our work the
Higgs term in the Lagrangian is replaced by three four-fermion
coupling terms between both $t$ and $b$ quarks, leading ultimately
to a massive bound state with the quantum numbers of the Higgs
boson.
We use a matrix formulation throughout, which straightforwardly
generalizes to $2n$ quarks.
There are massless neutral and charged bound states
with the quantum numbers of
Goldstone bosons of the broken SU(2)$_L\times{\rm U}(1)_R$ symmetry,
though the charged Goldstone boson bound state condition
highlights a failure of the bubble approximation when the lines of
the loop are two different particles. We give an {\em ad hoc} method for
dealing with this problem.
We start with no bias as to the relative
strengths of the three couplings for $t\bar t$, $b\bar b$, and
$b\bar t$. We assume that both the $t$ and $b$ quark fields acquire nonzero
vacuum expectation values, and concentrate on finding the general
results of adding a second flavor and $tb$ mixing terms to the interaction
Lagrangian. One set of our solutions agrees with Suzuki's extension, plus
we have striking new results not possible with only one flavor, and extendable
to any number of quark pairs.

We make the usual bubble or large $N_c$
expansion of the Nambu--Jona-Lasinio method, and study
the mass-gap equations and the
bound state conditions.
When there are  finite quark masses as solutions to the
mass-gap equations, we find a new
version of the fine-tuning constraint on the couplings $g_t, g_b, g_{tb}$.
In addition to the previous solution for the $g$'s for which
$g\Lambda ^2\sim$O(1), or $g\sim 1/\Lambda ^2$,
we find another solution with $g_tg_b - g_{tb}^2
\sim 1/\Lambda ^2$.
This expression is det {\bf G}, where {\bf G} is the matrix of
quark couplings.
This condition imposes an extra slightly broken
symmetry on the interaction Lagrangian ${\cal L}_{\rm int}$, a symmetry
which does not hold for the kinetic part of the Lagrangian and whose
significance is not yet clear. When det {\bf G} = 0, the symmetry
holds, the quarks are massless, and the Higgs and Goldstone boson
states are turned off.
When the symmetry is slightly broken,
the top and bottom masses are related by $m_t/m_b\approx
 g_b/g_{tb}$,
 and
the Higgs mass, in agreement with Suzuki, is given to leading order by
$m_H^2 = 4(m_t^4 + m_b^4)/(m_t^2 + m_b^2)$. Breaking the symmetry
corresponds to turning on one eigenvalue ${\rm G}_-$ of {\bf G} from 0 to
$1/C\Lambda ^2$, $C=N_c/8\pi ^2$, while the other eigenvalue ${\rm G}_+$
remains large, $(g/C)(1+m_t^2/m_b^2)$, as do all the couplings.
We thus do not need all couplings to be small to get $m_t,m_b,m_H\ll\Lambda $.

The key element in our method which allows generalization to any number
of quarks, and which led to recognizing the new solution, is the
observation that all the divergent self-energy integrals are given in
leading approximation by $C\Lambda ^2$, and appear in the mass-gap equations
multiplied by one of the couplings
$g$ times one of the quark masses $m$ in the self-energy loop.  This common
factor allows us to write the leading part of the mass-gap equations in
matrix form,
({\bf G}$-(1/C\Lambda ^2){\bf 1})${\bf m} =0,
where {\bf G} is the matrix of couplings and {\bf m}
is a vector of quark masses.  The usual fine-tuning condition is to take the
elements of {\bf G} of O($1/C\Lambda ^2$).  Our new solution comes from
expanding the couplings in series in $1/C\Lambda ^2$, with the leading term
in the series itself finite, not infinitesimal, as $\Lambda ^2$ approaches
infinity. Thus {\bf G} = ${\rm {\bf G}}_0 + {\rm O}(1/C\Lambda ^2)$,
with ${\rm {\bf G}}_0 \sim$ O(1).  The new fine-tuning constraint is then
det ${\rm {\bf G}}_0$ = 0.

If  ${\rm {\bf G}} = {\rm {\bf G}}_0$, the condition det ${\rm {\bf G}}_0$
= 0 corresponds to having a ``flat direction'' in the effective interactions
for composite $q\bar q$ states.  An eigenvector expansion shows that this is
the direction which becomes the Higgs and Goldstone bosons
when the symmetry is slightly broken,
{\bf G} = ${\rm {\bf G}}_0 + {\rm O}(1/C\Lambda ^2)$, and the quarks
acquire mass. This new solution may be particularly interesting for models
in which the symmetry is broken at a scale $\Lambda$ which is accessible,
{\it e.g.}, a few TeV, but is still enough greater than $m_t$ to make
$m_t/\Lambda$ negligible.  Then the dimensionless couplings
$g' = \Lambda ^2g$ would correspond to strong interactions.

In Section II we set up the matrix formulations of the Lagrangian,
mass-gap equations, and bound-state conditions. In section III we
show the new strong-coupling solution for the $g$'s, and in
section IV we discuss the new symmetry of ${\cal L}_{\rm int}$
 associated with this solution. Section V is a
brief summary of our results.

\section{COUPLING MATRIX FORMULATION OF THE INTERACTION
LAGRANGIAN, MASS-GAP EQUATIONS, AND BOUND-STATE EQUATIONS}

In this section, we find matrix expressions for all the key
elements of the multi-channel (two channels in this paper)
dynamical symmetry breaking problem. We write the interaction
Lagrangian ${\cal L}_{\rm int}$ for the most general $\rm
SU(2)_L\times U(1)_R$ invariant interaction between $t$ and $b$ quarks
with three couplings $g_t,g_b,g_{tb}$ in terms of the eigenvalues
and eigenvectors of a $2\times2$ coupling matrix {\bf G}, and also
split ${\cal L}_{\rm int}$ into pieces made up of neutral
combinations of either neutral or charged ``states" ($q\bar q$
bilinears).

Next we examine the mass-gap equations for this two-quark system
in the bubble approximation, that is, keeping only the leading
term in the $1/N_c$ expansion for the limit of large $N_c$. The
expressions for the quark self energies contain self-energy
integrals $I_t,I_b$ which require a cutoff mass scale $\Lambda $,
much larger than the quark masses $m_t,m_b$. There are two
mass-gap equations which must be simultaneously satisfied for the
quarks to have finite masses $m_t,m_b$; and these equations can be
written in terms of the $2\times2$ coupling matrix ${\bf G}$ and a
$2\times2$ diagonal matrix {\bf I} with $I_t$ and $I_b$ on the
diagonal, plus a 2-component mass vector {\bf m} with $m_t$ and
$m_b$ the components.

Finally we derive the one-loop or bubble approximation conditions
that there be three massless bound states corresponding to neutral
pseudoscalar and positive and negative charged Goldstone bosons, plus
one bound state which is a massive neutral scalar Higgs boson.
These bound-state conditions can also be expressed in terms of {\bf
G} and a $2\times2$ diagonal matrix {\bf J} with one-loop
integrals $J_t$ and $J_b$ as the diagonal elements.

The Goldstone theorem states that when the mass-gap equations are
satisfied, so that the original $\rm SU(2)_L\times U(1)_R$ is
dynamically broken, the three Goldstone bosons must exist. We
state the condition for satisfying the mass-gap equations as a
condition on the determinant of a matrix, and find that indeed we
need exactly the same condition on the same determinant to have a
neutral Goldstone boson. However, as has been noted elsewhere, the
bubble approximation is not quite good enough in the charged case
to yield a Goldstone boson. The problem is having a massive $t$
quark in the $b$ quark channels and {\em vice versa\/}. This problem is
neatly displayed in the matrix formulation, can be isolated, and
the further necessary approximation imposed.

The same determinant condition appears as the bound-state
condition for the massive scalar case, and in this case the piece
of the matrix which must equal zero is the mass constraint piece
expressing the mass of the Higgs $m_H$ in terms of $m_t$ and $m_b$.

The matrix formulations of ${\cal L}_{\rm int}$, the mass-gap
equations, and the bound-state conditions will make it very easy
in the following two sections to see the physical significance of
our new solution for the $g_i$ ($i=t,b,tb$) and our new symmetry
of ${\cal L}_{\rm int}$.

\subsection{Lagrangian}

Our interaction Lagrangian is
%
% equation
%
\begin{equation}
\label{L1}
	{\cal L}_{\rm int}=g_t\bar\psi _Lt_R\bar t_R\psi _L
	+g_b\bar\psi _Lb_R\bar b_R\psi _L
	+g_{tb}\left( \bar\psi _Lb_R\bar t_{R}^c\tilde\psi _L
	+\overline{\tilde\psi _L}t_{R}^c\bar b_R\psi _L \right),
\end{equation}
where all $L$'s are $\rm SU(2)$ doublets and all $R$'s are SU(2)
singlets, and
%
% equation
%
\begin{equation}
\label{SU2}
        \psi_L=\left(
\begin{array}{c}
	t_L\\
	b_L
\end{array}
 \right)\quad {\rm and}\quad \tilde\psi _L=\left(
 \begin{array}{c}
 	-b_{L}^c\\
 	t_{L}^c
 \end{array}
  \right),
\end{equation}
\begin{equation}
	t_L^c\equiv (t_L)^c=\frac{1+\gamma ^5}{2}t^c=\frac{1+\gamma
	^5}{2}C\gamma ^0(t^\dagger)^T\label{tL}.
\end{equation}
In each bilinear factor there is an implied summation over color indices,
{\em i.e.}, $\bar\psi _Lt_R=\sum_{a}^{}\bar\psi _L^at_R^a$.
What we call $g_t$,
  the coupling associated with the $t_R\bar t_R$ term, is the only
  coupling in the single (top) quark condensate
  model. The Lagrangian can be expressed entirely in terms of
  $L,R$ spinors or in terms of $\bf 1$ and $\gamma ^5$, and can be
  separated into neutral-neutral and charged-charged interactions.

  It is convenient to define a $2\times2$ coupling matrix $\bf G$,
  %
  % equation
  %
  \begin{equation}
  \label{G1}
  	{\bf G}=\left(
  	\begin{array}{cc}
  		g_t &-g_{tb}\\
  		-g_{tb} &g_b
  	\end{array}
  	 \right),
  \end{equation}
  together with state vectors ${\bf x}$ and ${\bf x}'$,
  respectively representing neutral and charged states,
  %
  % equation
  %
  \begin{equation}
  \label{x1}
  \begin{array}{c@{\hspace{3em}}c@{\hspace{3em}}c@{\hspace{3em}}c}
  	{\bf x}=\left(
  	\begin{array}{c}
  		x_1\\
  		x_2
  	\end{array}
  	 \right), & {\bf x}^\dagger=\left( x_1^\dagger,\   x_2^\dagger
  	  \right), & {\bf x}'=\left(
  	  \begin{array}{c}
  	  	x_1'\\
  	  	x_2'
  	  \end{array}
  	   \right), & {{\bf x}'}^\dagger =\left( {x_1'}^\dagger,\
  	   {x_2'}^\dagger \right)
  \end{array}
  ,
  \end{equation}
where
%
% equation
%
\begin{equation}
\label{x2}
\begin{array}{c@{\hspace{3em}}c@{\hspace{3em}}c@{\hspace{3em}}c@{\hspace{3em}}}
	x_1=\bar t_Lt_R, & x_2=\bar b_Rb_L, & x_1'=\bar b_Lt_R, & x_2'=-\bar
	b_R t_L,\\
	x_1^\dagger = \bar t_Rt_L, & x_2^\dagger=\bar b_Lb_R, &
	{x_1'}^\dagger=\bar t_Rb_L, & {x_2'}^\dagger=-\bar t_L b_R.
\end{array}
\end{equation}

${\cal L}_{\rm int}$ can then be written  in terms of ${\bf G}$ as
%
% equation
%
\begin{equation}
\label{L2}
	{\cal L}_{\rm int}={\bf x}^\dagger{\bf Gx}+{{\bf x}'}^\dagger{\bf
	Gx}'.
\end{equation}
In this form, the interaction Lagrangian lends itself to
diagonalization in terms of the eigenvalues and eigenvectors of
${\bf G}$. Let the eigenvalues be $G_+$ and $G_-$,
%
% equation
%
\begin{equation}
\label{eval1}
	G_\pm=\frac{g_t+g_b}{2}\pm\frac{1}{2}\sqrt{\left( g_t+g_b
	\right)^2-4\left( g_tg_b-g_{tb}^2 \right)},
\end{equation}
with eigenvectors $\mbox{\boldmath$\xi $}_+$ and
$\mbox{\boldmath$\xi $}_-$,
%
% equation
%
\begin{eqnarray}
\label{evec1}
	\mbox{\boldmath$\xi $}_+ &=& \left(
	\begin{array}{c}
		a_+\\
		b_+
	\end{array}
	 \right)=N \left(
	 \begin{array}{c}
	 	1\\
	 	-g_{tb}/\left(g_t-G_-\right)
	 \end{array}
	  \right)
	  =N\left(
	  \begin{array}{c}
	  	1 \\ a
	  \end{array}
	    \right),
	   \nonumber \\
	  \mbox{\boldmath$\xi $}_- &=& \left(
	  \begin{array}{c}
	  	a_-\\
	  	b_-
	  \end{array}
	   \right)=N \left(
	   \begin{array}{c}
	   	-g_{tb}/\left(g_t-G_-\right)\\
	   	-1
	   \end{array}
	    \right)=N\left(
	    \begin{array}{c}
	    	a\\ -1
	    \end{array}
	     \right),
	    \nonumber \\
            N &=& \frac{1}{\sqrt{ 1+g_{tb}^2/
	\left( g_t-G_-\right)^2}}=\frac{1}{\sqrt{1+a^2}}.
\end{eqnarray}
We take the matrix {\bf G} to be positive, semi-definite, so that
the eigenvalues are either zero or positive, and the interaction
is attractive. If {\bf G} should be negative semi-definite, we can
simply switch $G_\pm$.
The interaction Lagrangian can be written using the eigenvalues
and eigenvectors as a sum of two terms which involve only the
neutral states $\bf x$ and two which involve only the charged states
$\bf x'$,
%
% equation
%
\begin{equation}
\label{L3}
	{\cal L}_{\rm int}=G_+\left( {\bf x}^\dagger\mbox{\boldmath$\xi
	$}_+ \right)\left( \mbox{\boldmath$\xi $}_+^\dagger{\bf x} \right)+
	G_-\left( {\bf x}^\dagger \mbox{\boldmath$\xi $}_-\right)\left(
	\mbox{\boldmath$\xi $}_-^\dagger{\bf x} \right)
	+G_+\left( {{\bf x}'}^\dagger\mbox{\boldmath$\xi $}_+ \right)
	\left( \mbox{\boldmath$\xi $}_+^\dagger{{\bf x}'} \right)
	+G_-\left( {{\bf x}'}^\dagger\mbox{\boldmath$\xi $}_- \right)
	\left( \mbox{\boldmath$\xi $}_-^\dagger{\bf x}' \right).
\end{equation}
Some simple identities, which will be useful when we return to this form
of ${\cal L}_{\rm int}$ later, with particular interest in the
case that the eigenvalue $G_-=0$, are
%
% eqnarray
%
\begin{eqnarray}
\label{eval2}
	G_+G_- = g_tg_b-g_{tb}^2 = {\rm det}{\bf G},\qquad\quad
	G_++G_- &=& g_t+g_b,\qquad\quad
	g_t-G_- = G_+-g_b,\nonumber \\
	\mbox{\boldmath$\xi$} _\pm^\dagger
	\mbox{\boldmath$\xi$} _\pm = a_\pm^2+b_\pm^2=1,\qquad\quad
	\mbox{\boldmath$\xi$} _\pm^\dagger
	\mbox{\boldmath$\xi$} _\mp &=& a_+a_-+b_+b_-=0.
\end{eqnarray}

\subsection{Mass-gap equations}

The Lagrangian of Eqs.\ (\ref{L1}), (\ref{L2}) or (\ref{L3}) describes
massless fermions unless certain mass-gap equations are satisfied.
These come from equating the self-energy $\Sigma _q$ of each quark
to its mass $m_q$, and, if satisfied, lead to a dynamical breaking
of the $\rm SU(2)_L\times U(1)_R$ symmetry.
We will treat the mass-gap equations for this two-quark system by
looking at the bubble approximation, or leading term in the ${\rm 1/N_c}$
expansion for the limit of large ${\rm N_c}$.
 It is
sufficient to sketch the calculation of $\Sigma _b$, the
self-energy of the $b$ quark; for
example, the $g_b$ terms in the Lagrangian are
%
% eqnarray
%
\begin{eqnarray}
\label{Lb}
	{\cal L}_{\rm int}^{(g_b)} &=& g_b\left( \bar t_Lb_R\bar
	b_Rt_L+\bar b_Lb_R\bar b_Rb_L \right)\nonumber \\
	 &=& \frac{1}{4}g_b\left( \bar tb\bar bt-\bar t\gamma ^5b\bar b\gamma
	 ^5t+\bar bb\bar bb-\bar b\gamma ^5b\bar b\gamma ^5b +\bar b t
	 \bar t \gamma ^5b -\bar t b\bar b\gamma ^5t\right).
\end{eqnarray}
The only contribution to $\Sigma _b$ from the $g_b$ terms in the
bubble approximation is
%
% eqnarray
%
\begin{eqnarray}
\label{Siggb}
	\Sigma _b^{(g_b)}
	 &=& g_bm_bI_b\nonumber\\
	 &=& g_bm_b\frac{2N_c}{(2\pi )^4}\int_{\Lambda
	 }d^4p\frac{i}{p^2-m_b^2}\nonumber\\
	&=& g_bm_b\frac{2N_c}{(2\pi
	)^4}\int_{0}^{\Lambda }\frac{d^4p_E}{p_E^2+m_b^2},
\end{eqnarray}
with $m_b$ the physical mass of the $b$ quark and $p_E$ a
Euclidean 4-momentum. The integral $I_b$ is typical of integrals
in self-energy expressions,
%
% equation
%
\begin{equation}
\label{Ii}
	I_i=\frac{2N_c}{(2\pi )^4}\int_{0}^{\Lambda
	}\frac{d^4p_E}{p_E^2+m_i^2},
\end{equation}
all defined with the same cutoff $\Lambda $, which is assumed to be much
larger than the quarks' masses. Similarly to the $g_b$ case, the
$g_t$ term gives no contribution to $\Sigma _b$ and the $g_{tb}$
contribution is
%
% eqnarray
%
\begin{eqnarray}
\label{Sigtb}
	\Sigma _b^{(g_{tb})} &=& -g_{tb}m_tI_t\nonumber \\
	 &=&
	 -g_{tb}m_t\frac{2N_c}{(2\pi
	)^4}\int_{0}^{\Lambda }\frac{d^4p_E}{p_E^2+m_t^2},
\end{eqnarray}
with $m_t$ the physical mass of the $t$ quark.

The total contribution to $\Sigma _b$ is $\Sigma _b^{(g_b)}+\Sigma
_b^{(g_{tb})}$. The physical mass $m_b$ of the $b$ quark is
the solution of the mass-gap equation
%
% eqnarray
%
\begin{eqnarray}
\label{mb}
	m_b  &=& \left[\Sigma _b^{(g_b)}\left( p^2 \right)+\Sigma
	_b^{(g_{tb})}\left( p^2 \right)\right]_{p^2=m_b^2}\nonumber \\
	 &=& g_bm_bI_b-g_{tb}m_tI_t.
\end{eqnarray}
The expression for the $t$ quark physical mass $m_t$ is derived in
a similar way and is the solution of the mass-gap equation,
%
% equation
%
\begin{equation}
\label{mt}
	m_t=g_tm_tI_t-g_{tb}m_bI_b.
\end{equation}
The mass-gap Equations (\ref{mb}) and (\ref{mt}) must be satisfied
simultaneously for
the $b$ and $t$ quarks to have finite mass.
The two equations  can be written in terms of a
$2\times2$ matrix and a 2-component mass vector ${\bf m}=\left(
m_t\atop m_b \right)$ as
%
% equation
%
\begin{equation}
\label{G-1,1}
	\left(
	\begin{array}{c@{\hspace{1em}}c}
		g_tI_t-1 & -g_{tb}I_b\\
		-g_{tb}I_t & g_bI_b-1
	\end{array}
	 \right)\left(
	 \begin{array}{c}
	 	m_t\\ m_b
	 \end{array}
	  \right) = 0.
\end{equation}
However, the dependence of the $I$'s on the $m$'s makes these
nonlinear equations in the masses.

We define a diagonal $2\times2$ matrix {\bf I},
\begin{equation}\label{I,1}
	{\bf I}=\left(
	\begin{array}{cc}
		I_t & 0\\
		0 & I_b
	\end{array}
	 \right),
\end{equation}
so that the mass-gap equations are
\begin{equation}
	\left( {\bf GI-1} \right){\bf m}=0,\label{GI-1,1}
\end{equation}
and the condition for a finite solution $m_t,m_b$ is
\begin{equation}
	{\rm det}\left( {\bf GI-1} \right)=0\label{detGI-1}
\end{equation}
when {\bf I} is evaluated in terms of the physical masses.

The integrals $I_t$ and $I_b$ are both quadratically divergent,
and as long as $\Lambda ^2/m^2 \gg 1$,
%
% eqnarray
%
\begin{eqnarray}
\label{Is1}
	I_t &=& C\left( \Lambda ^2-m_t^2\ln\frac{\Lambda ^2}{m_t^2}
	+{\rm O}\left( \frac{m_t^4}{\Lambda ^2} \right)\right),\nonumber\\
	I_b &=& C\left( \Lambda ^2-m_b^2\ln\frac{\Lambda ^2}{m_b^2}
	+{\rm O}\left( \frac{m_b^4}{\Lambda ^2} \right)\right),
\end{eqnarray}
where $C=N_c/8\pi ^2$.  For many purposes the slight
difference in magnitude between $I_t$ and $I_b$ can be ignored
relative to their overall magnitude, in which case
%
% equation
%
\begin{equation}
\label{Is2}
	I_t\simeq I_b\simeq C\Lambda ^2.
\end{equation}
To this order, the mass-gap Eqs.\ (\ref{G-1,1}) then become
linear, homogeneous equations in the masses,
%
% equation
%
\begin{eqnarray*}
	\left(
	\begin{array}{c@{\hspace{1em}}c}
		g_t-\frac{1}{C\Lambda ^2} & -g_{tb}\\
		-g_{tb} & g_b-\frac{1}{C\Lambda ^2}
	\end{array}
	 \right)\left(
	 \begin{array}{c}
	 	m_t\\ m_b
	 \end{array}
	  \right) = 0,
\end{eqnarray*}
or
%
% equation
%
\begin{equation}
\label{G-1,2}
	\left( {\bf G}-\frac{1}{C\Lambda ^2}{\bf 1} \right){\bf m} = 0.
\end{equation}
One way to satisfy this (leading order)
condition is for one of the eigenvalues of
${\bf G}$ to be equal to $1/C\Lambda ^2$, while at the same time
the mass vector ${\bf m}$ is proportional to the corresponding
eigenvector. We will show in Sec.\ III, see
Eqs.\ (\ref{Aev1}) and (\ref{Bev1}),
that this is a natural solution for
the eigenvalue we called $G_-$ in Eq.\ (\ref{eval1}),
%
% equation
%
\begin{equation}
\label{G-,1}
	G_- = \frac{1}{C\Lambda ^2}\quad\mbox{and}\quad {\bf
	m}\propto\mbox{\boldmath$\xi $}_-.
\end{equation}

\subsection{Bound-state conditions}

We now set up the (dynamical symmetry-breaking) bound-state
conditions for
three massless bound states corresponding to neutral pseudoscalar
and positive and negative charged Goldstone bosons, plus a bound
state which is a massive neutral scalar Higgs boson. In the simple
sum of one-loop or bubble graphs we are considering, there is only
an $s$-channel, no $t$- or $u$- channels, and eigenstates scatter
to eigenstates, ${\bf 1\rightarrow1},\gamma ^5\rightarrow\gamma
^5,LR\rightarrow LR$. Diagonalizing the interaction to find the
eigenstates amounts to diagonalizing the $s$-channel scattering
matrix.

The scattering matrix can be written as
\begin{eqnarray}\label{GJG}
	\frac{1}{2}\left[ {\bf G+GJG+GJGJG+\cdots} \right] &=&
	\frac{1}{2}{\bf G}\frac{1}{{\bf 1-JG}}\nonumber\\
	 &=& \frac{1}{2}{\bf G}\frac{1}{{\bf G-GJG}}{\bf G}\nonumber\\
	  &=& \frac{1}{2}{\bf G}\left[ {\rm cofac}\left( {\bf 1-JG}
	  \right) \right]^T\frac{1}{{\rm det}\left( {\bf 1-JG} \right)},
\end{eqnarray}
where
\begin{equation}
	\label{J}
	{\bf J}=\left(
	\begin{array}{cc}
		J_t & 0\\ 0 & J_b
	\end{array}
	 \right),
\end{equation}
and $J_i$ is a one-loop function (integral). The condition for a
bound state is that the scattering matrix have a pole, {\em i.e.}, that
the denominator in the last form of Eq.\ (\ref{GJG}) vanish,
\begin{equation}
	\label{det1-JG}
	{\rm det}\left( {\bf 1-JG} \right)=0.
\end{equation}
Since {\bf G} is symmetric and {\bf J} is diagonal, we have
\begin{equation}
	\label{detGJ-1,1}
	{\rm det}\left( {\bf GJ-1} \right)=0.
\end{equation}
We will show that the one-loop integrals $J_i$ can be expressed in
terms of the self-energy integrals $I_i$ as
\begin{equation}
	\label{J1delta}
	J_i=I_i+\Delta J_i,
\end{equation}
so the bound-state condition becomes
\begin{equation}
	\label{detGJ-1,2}
	{\rm det}\left( {\bf GI+G}\Delta {\bf J-1} \right)=0.
\end{equation}

If the mass-gap equations are satisfied, then Eq.\
(\ref{detGI-1}), ${\rm det}\left( {\bf GI-1} \right)=0$, already
holds, and we can use straightforward matrix manipulations to find
the conditions for bound states, which must exist by the Goldstone
Theorem and Nambu--Jona-Lasinio construction. Our procedure
follows, and is easily generalized to  $2n\times2n$, not only
$2\times2$, matrices.

When ${\rm det}\left( {\bf GI-1} \right)=0$, in its diagonal form
the matrix $\left( {\bf GI-1} \right)$ must have one zero
eigenvalue. Other eigenvalues must be of leading order $GC\Lambda
^2$, where $G$ is a ``large" eigenvalue of {\bf G}. For our $2\times2$
case, we choose the ``minus" eigenvalue $G_-$ of $\left( {\bf GI-1}
\right)$ to be zero, and call the corresponding eigenstate
$\mbox{\boldmath$\xi $}'_-$.

Now we write the extra piece ${\bf G}\Delta{\bf J}$ of $\left( {\bf
GJ-1} \right)$ in the $-,+$ basis, so the matrix elements are
labeled $--,-+,+-,++$. The full matrix $\left( {\bf GJ-1} \right)$
is thus these four matrix elements, plus the nonzero eigenvalue of
({\bf GI-1}) in the $++ $ position,
\begin{equation}
	\left( {\bf GJ-1} \right)=\left(
	\begin{array}{cc}
	\mbox{\boldmath$xi$}'^T_-{\bf G}\Delta{\bf J}\mbox{\boldmath$xi$}'_- &
	\mbox{\boldmath$xi$}'^T_-{\bf G}\Delta{\bf J}\mbox{\boldmath$xi$}'_+\\
	 \mbox{\boldmath$xi$}'^T_+{\bf G}\Delta{\bf J}\mbox{\boldmath$xi$}_- &
\mbox{\boldmath$xi$}^T_+{\bf G}\Delta{\bf J}\mbox{\boldmath$xi$}'_+ +
	 G_+C\Lambda ^2-1\end{array}\right)
\end{equation}
Clearly for the determinant to be zero in leading approximation,
$\mbox{\boldmath$xi$}'^T_-{\bf G}\Delta{\bf J}\mbox{\boldmath$xi$}'_-$ must be
zero, since $G_+C\Lambda
^2$ is so large. (In the $2n\times2n$ case, there would be $2n-1$
large eigenvalues of ${\bf GI-1}$ on the diagonal.) All other
terms, as we will show explicitly later, are at most O(ln$\Lambda
^2$).

Our bound-state condition is now reduced to setting one matrix
element equal to zero, and we now make some approximations, all
good to ${\rm O}(\ln\Lambda ^2/\Lambda ^2)$.
First, we consider the relation of $\mbox{\boldmath$xi$}'_\pm$, the
eigenvectors of $({\bf GI-1})$, to $\mbox{\boldmath$xi$}_\pm$, the eigenvectors
of ${\bf G}$. The matrix $\bf I$ is diagonal,
\begin{equation}
	\label{I,2}
	{\bf I}=\left(
	\begin{array}{cc}
		C\Lambda ^2-m_t^2\frac{\Lambda ^2}{m_t^2}+{\rm O}\left(
		\frac{m_t^4}{\Lambda ^2} \right) & 0\\
		0 & C\Lambda ^2-m_b^2\ln\frac{\Lambda ^2}{m_b^2}+{\rm O}\left(
		\frac{m_t^4}{\Lambda ^2} \right)
	\end{array}
	 \right),
\end{equation}
and to leading order is just ${\bf I}\simeq C\Lambda ^2{\bf 1}$.
Thus to leading order the same eigenvectors $\mbox{\boldmath$xi$}_\pm$
diagonalize {\bf GI} as diagonalize {\bf G} alone. Since ${\bf
\Delta J}$ is diagonal, our bound-state condition is now
\begin{equation}
	\label{delta--}
	\mbox{\boldmath$xi$}^T_-{\bf \Delta
JG}\mbox{\boldmath$xi$}_-=\mbox{\boldmath$xi$}^T_-{\bf \Delta
	J}G_-\mbox{\boldmath$xi$}_-=0.
\end{equation}
The $G_-$ factors out, and our generic condition for a bound state
in the one-loop approximation, when the mass-gap equations are
satisfied, in terms of the components of $\mbox{\boldmath$xi$}_-$, is
\begin{equation}
	a_-^2\Delta J_t+b_-^2\Delta J_b=0,
\end{equation}
where $a_-$ and $b_-$ are defined in Eq. (\ref{evec1}).
Next we show the specific cases.

The neutral scalar channel terms in ${\cal L}_{\rm int}$ which
contribute to the bubble expansion scattering matrix are
\begin{equation}
\label{Lneus}
	{\cal L}_{\rm int}^{\rm neu\ s}=\frac{1}{4}g_t\bar tt\bar
	tt+\frac{1}{4}g_b\bar bb\bar bb-\frac{1}{2}g_{tb}\bar tt\bar bb .
\end{equation}
The one-loop integral $J_i$ is $(i=t,b)$,
\begin{eqnarray}\label{JintH}
	J_i &=& -\frac{i}{2}\frac{N_c}{\left( 2\pi
	\right)^4}\int d^4k\,Tr\frac{i}{\left( \not p+\not k
	\right)-m_i}\,\frac{i}{\not k-m_i}\nonumber\\
	 &=& i\frac{2N_c}{\left( 2\pi  \right)^4}\int
	 d^4k\,\frac{k\cdot(p+k)+m_i^2}{\left[ (p+k)^2-m_i^2
	 \right]\left[ k^2-m_i^2 \right]}\nonumber \\
	 &=& i\frac{2N_c}{(2\pi )^4}\left\{ \int
	 d^4k\frac{1}{k^2-m_i^2}-\frac{1}{2}(p^2-4m_i^2)\int
	 d^4k\frac{1}{\left[ (p+k)^2-m_i^2 \right]\left[ k^2-m_i^2
	 \right]} \right\}.
\end{eqnarray}
We recognize the first integral as $I_i$, see Eq.\ (\ref{Ii}),
so that we can write $J_i$ as
\begin{eqnarray}\label{JhasI,1}
	J_i &=& I_i+\frac{C}{2}(p^2-4m_i^2)\nonumber\\
	    &=& I_i+\Delta J_i.
\end{eqnarray}
We neglect the difference between $\ln(\Lambda ^2/m_t^2)$ and
$\ln(\Lambda ^2/m_b^2)$, and find that to have a neutral scalar
bound state, we must have
\begin{equation}
	a_-^2(p^2-4m_t^2)+b_-^2(p^2-4m_b^2)=0.\label{polesn}
\end{equation}
In the next section, we will find the values of $G_\pm$,
$\mbox{\boldmath$xi$}_\pm$ which allow the mass-gap equation to hold. For now
we invoke those solutions, where $a_-^2=m_t^2/(m_t^2+m_b^2)$ and
$b_-^2=m_b^2(m_t^2+m_b^2)$.  This gives us the (leading order)
constraint on the neutral scalar bound state, identified as the
Higgs boson,
\begin{equation}
\label{mH1}
	p^2=m_H^2=\frac{4(m_t^4+m_b^4)}{(m_t^2+m_b^2)},
\end{equation}
in agreement with Suzuki \cite{s}.
Because we know that $m_t^2\agt1200m_b^2$, see Ref.\ \cite{cdf},
this places the mass of
the Higgs very near $m_H=2m_t$,
%
% equation
%
\begin{equation}
\label{mH2}
	m_H=2m_t\left[ 1-\frac{1}{2}\frac{m_b^2}{m_t^2}+{\rm
	O}\left( \frac{m_b^4}{m_t^4} \right) \right].
\end{equation}
Thus $m_H$  is barely changed from the result of Bardeen {\em et
al.\/} \cite{bhl}, $m_H=2m_t$, by the inclusion of a second quark vacuum
expectation value (vev).

The pseudoscalar neutral channel terms in ${\cal L}_{\rm int}$
which contribute to the scattering matrix are
\begin{equation}
	{\cal L}_{\rm int}^{\rm neu\ ps}=-\frac{1}{4}g_t\bar t\gamma
	^5t\bar t\gamma ^5t-\frac{1}{4}g_b\bar b\gamma ^5b\bar b\gamma
	^5b-\frac{1}{2}g_{tb}\bar t\gamma ^5t\bar b\gamma ^5b.\label{Lneups}
\end{equation}
The one-loop integral $J_i$ is
\begin{eqnarray}\label{JintGO}
	J_i &=& -\frac{i}{2}\frac{N_c}{(2\pi )^4}\int
	d^4k\,Tr\frac{i}{\not k-m_i}\gamma ^5\frac{i}{\not p+\not
	k-m_i}\gamma ^5\nonumber \\
	 &=& -i\frac{2N_c}{(2\pi )^4}\int
	 d^4k\,\frac{k\cdot(p+k)-m_i^2}{\left[ (p+k)^2-m_i^2
	 \right][k^2-m_i^2]},
\end{eqnarray}
or
\begin{eqnarray}\label{JhasI,2}
	J_i &=& -I_i-\frac{C}{2}p^2\left[ \ln\frac{\Lambda
	^2}{m_i^2}+\cdots \right]\nonumber \\
	 &=& -I_i-\Delta J_i.
\end{eqnarray}
The signs of the $g_t,g_b$ couplings in Eq.\ (\ref{Lneups}) are
opposite to Eq.\ (\ref{Lneus}), which means that the coupling matrix is
not ${\bf G}$, but ${\tilde {\bf G}}$,
\begin{eqnarray}\label{Gtilde}
	\tilde {\bf G}=\left(
	\begin{array}{cc}
		1&0\\ 0&-1
	\end{array}
	 \right){\bf G}\left(
	 \begin{array}{cc}
	 	1&0\\ 0&-1
	 \end{array}\right).
\end{eqnarray}
It is not obvious that
${\rm
det}\left( {\bf \tilde GI+\tilde G\Delta J-1} \right)=0$.
However, it is, and furthermore we find after some manipulations
that the bound-state condition we have reduces to ${\rm det}({\bf
GI+G\Delta J-1})=0$, thanks to the commutativity of diagonal matrices
and the properties of determinants,
\begin{eqnarray}\label{det-GI}
	{\rm det}({\bf \tilde GI+\tilde G\Delta J-1})
	&=& {\rm det}\left\{\left(
	      \begin{array}{cc}
	      	1&0\\ 0&-1
	      \end{array}
	       \right)({\bf GI}+{\bf G\Delta J}-
	{\bf 1})\left(
	         \begin{array}{cc}
	         	1&0\\ 0&-1
	         \end{array}
	          \right)\right\}\nonumber \\
	           &=& {\rm det}({\bf GI+G\Delta J-1})=0.
\end{eqnarray}
This construction generalizes to the $2n\times2n$ matrices encountered with
$n$ generations of quarks.

 In this case, then, $\Delta J\propto p^2$, and the bound state
 constraint reduces to $p^2=0$. This is the massless pseudoscalar
 neutral Goldstone boson.

 The charged channel terms which contribute to the scattering matrix,
 left in terms of L,R in ${\cal L}_{\rm int}$, are
 \begin{eqnarray}\label{Lch}
 	{\cal L}_{\rm int}^{\rm ch} &=& g_t\bar b_Lt_R\bar
 	t_Rb_L+g_b\bar t_Lb_R\bar b_Rt_L\nonumber \\
 	&&+g_{tb}\bar t_Lb_R\bar b_Lt_R+g_{tb}\bar b_Rt_L\bar t_Rb_L.
 \end{eqnarray}
 Thus all the diagrams involve $b\leftrightarrow t$ loops, and the
 natural inclination is to assume
 $m_t\not=0,m_b\not=0,m_t\not=m_b$. We write as illustration the
 $J^\pm$ integral for a $\bar bt$ diagram,
 \begin{eqnarray}\label{J+-1}
 	J^\pm &=& -i\frac{N_c}{(2\pi )^4}\int d^4k\,Tr\frac{i}{\not
 	k-m_b}\,\frac{1\pm\gamma ^5}{2}\,\frac{i}{\not p+\not
 	k-m_t}\,\frac{1\mp\gamma ^5}{2}\nonumber \\
 	 &=& \frac{2N_c}{(2\pi )^4}\int
 	 d^4k\,\frac{k\cdot(p+k)}{(k^2-m_b^2)((p+k)^2-m_t^2)}.
 \end{eqnarray}
 We can express this in terms of $I_t$ and $I_b$ as
 \begin{equation}
 	\label{J+-2}
 	J^\pm=I_t+I_b-\frac{C}{2}(p^2-m_t^2-m_b^2)\int d^4
 	k\,\frac{1}{(k^2-m_b^2)[(p+k)^2-m_t^2]},
 \end{equation}
 where the integral is logarithmically divergent, or in terms of
 $I_t$ alone as
 \begin{equation}
 	\label{J+-3}
 	J^\pm=I_t+p^2I_1+m_b^2I_2,
 \end{equation}
 where $I_1$ and $I_2$ are both logarithmically divergent, and
 both have the denominator containing both masses.

 Alarmingly, in this one-loop approximation for the multichannel
 case, the Goldstone Theorem fails: there should be a $p^2=0$
 (massless) charged Goldstone boson. Clearly, dynamical symmetry
 breaking is more subtle than a bubble approximation. By putting
 in finite unequal masses we have already broken the symmetry by
 hand. This problem was not noted by Suzuki \cite{s} and was noted
 but not discussed by Pham \cite{ph}. There is an {\em ad hoc\/} remedy:
 set $m_t\not=0,m_b=0$ in loops with coupling $g_t$; set
 $m_b\not=0,m_t=0$ in loops with coupling $g_b$; and do one and then the
 other separately in the two off-diagonal terms in the matrix {\bf GJ}
which correspond to loops with coupling $g_{tb}$. We do not give these
 details here. The result is that ${\bf JG=IG}+{\rm O}(p^2)$, so
 as in the neutral pseudoscalar case, there is a $p^2=0$ massless
 charged bound state.

 \section{Two solutions for couplings which give finite quark masses}

 We reverse the usual approach to solving the mass-gap equations,
 Eq.\ (\ref{G-1,1}). Instead of assuming that the couplings in
 {\bf G} are known and solving for allowed m's, we assume that
 $m_t,m_b$ exist, are finite, and are unequal, $\Lambda \gg
 m_t>m_b$. This of course makes sense experimentally. Then we look
 for constraints on the $g$'s which allow solutions to Eq.\
 (\ref{G-1,1}). We find two sets of solutions. The first set is
 known, see Suzuki \cite{s}, and is the generalization of the
 Bardeen, Hill, and Lindner fine-tuning constraint on the top
 quark coupling of \cite{bhl}. It is found by approximating
 $I_i\simeq C\Lambda ^2$ and solving Eq.\ (\ref{G-1,2}). The
 second solution, new in this paper, is found by expanding and
 matching terms order-by-order in the mass-gap equations.

 To get the first set of $g$'s we assume that all the couplings are
 of magnitude $1/C\Lambda ^2$,
 \begin{equation}
 	\label{gi1}
 	g_i=\frac{\hat g_i}{C\Lambda ^2},\quad\quad {\bf G}=\frac{\hat{\bf
 	G}}{C\Lambda ^2}.
 \end{equation}
 This condition makes all the $gI$ terms in Eq.\ (\ref{G-1,1})
 finite in the limit $\Lambda ^2\rightarrow\infty$. These $g$'s
 are a (leading order) solution of Eq.\ (\ref{G-1,2}) provided
 \begin{equation}
 	\label{detGhat1}
 	{\rm det}(\hat{\bf G}-{\bf 1})=0
 \end{equation}
 up to terms of O$(1/\Lambda ^2)$, a constraint on the coupling
 matrix which has apparently not been noted before. In this
 solution, the ``fine-tuning" is the condition $g\Lambda
 ^2\sim{\rm O}(1)$.

 The second set of $g$'s comes from allowing $g\Lambda ^2\gg1$,
 which allows a natural Higgs mass $m_H\ll\Lambda $. In this case
 the $g_i$ are of O(1), plus corrections of O$(1/\Lambda ^2)$, and
 the condition
 \begin{equation}
 	\label{detG1}
 	{\rm det}{\bf G}=(g_tg_b-g_{tb}^2)=0
 \end{equation}
 must hold up to terms of O$(1/\Lambda ^2)$. This solution
 involves a new symmetry as the eigenvalue $G_-\rightarrow0$.

 We note that for both sets of solutions, corrections of
 O$(1/\Lambda ^2)$ are essential to get nontrivial solutions to
 the mass-gap equations.

 The solution for the $g_i$ noted in Eq.\ (\ref{gi1}) is analogous
 to the fine-tuning solutions of \cite{bhl} and \cite{s} with the
 couplings set very small to give finite fermion masses,
 \begin{equation}\label{Bgi1}
 	g_t=\frac{\hat g_t}{C\Lambda ^2},\quad\quad g_b=\frac{\hat
 	g_b}{C\Lambda ^2},\quad\quad g_{tb}=\frac{\hat g_{tb}}{C\Lambda ^2}.
\end{equation}
We can express the $\hat g_i$ which allow positive, finite masses
$m_t,m_b$ in terms of a single constant $\hat g$ and the two masses,
\begin{equation}
\label{Bgi2}
	\hat g_t=\hat g,\quad\quad \hat g_b=1+\frac{m_t^2}{m_b^2}(\hat g-1),
	\quad\quad
	\hat g_{tb}=\frac{m_t}{m_b}(\hat g-1).
\end{equation}
The determinant of {\bf G} is
\begin{equation}\label{BdetG}
	{\rm det}{\bf G}=\frac{1}{C^2\Lambda ^4}\left[
	\hat g+\frac{m_t^2}{m_b^2}(\hat g-1) \right].
\end{equation}
The eigenvalues and eigenvectors of {\bf G} are
\begin{eqnarray}\label{Bev1}
	G_+ &=& \frac{1}{C\Lambda ^2}\left[\hat g+\frac{m_t^2}{m_b^2}(\hat g-1)
	\right],\nonumber \\
	G_- &=& \frac{1}{C\Lambda ^2},\nonumber \\
	\mbox{\boldmath$xi$}_+  &=& \left(
	\begin{array}{c}
		a_+\\ b_+
	\end{array}
	 \right)=N\left(
	 \begin{array}{c}
	 	1\\ m_t/m_b
	 \end{array}
	  \right)=N\left(
	  \begin{array}{c}
	  	1\\ a
	  \end{array}
	   \right),\nonumber \\
	  \mbox{\boldmath$xi$}_i &=& \left(
	  \begin{array}{c}
	  	a_-\\ b_-
	  \end{array}
	   \right)=N\left(
	   \begin{array}{c}
	   	m_t/m_b\\ -1
	   \end{array}
	    \right)=N\left(
	    \begin{array}{c}
	    	a\\ -1
	    \end{array}
	     \right),\nonumber \\
	    N &=&
	    \frac{1}{\sqrt{1+m_t^2/m_b^2}}=\frac{1}{\sqrt{1+a^2}}.
\end{eqnarray}
The mass ratio $m_t/m_b$ in terms of $\hat g$'s is
\begin{equation}
	\label{Bmtmb1}
	\frac{m_t}{m_b}=\frac{\hat g_b-1}{\hat g_{tb}}.
\end{equation}
This solution corresponds to a small perturbation by the mixing $g_{tb}$
of the solution of Ref. \cite{bhl} for two {\em independent}
$\langle \bar tt \rangle$, $\langle \bar bb \rangle$ vevs with
$g_t \approx g_b \approx 1/C\Lambda^2$, $\hat g_t \approx \hat g_b \approx 1$.

To get the new solution for the $g$'s, we solve Eq.\ (\ref{G-1,1})
explicitly, with $g_{tb}$ real and $m_t,m_b$ free
parameters. We find
\begin{eqnarray}\label{gnew1}
	g_{tb} &=&
	\frac{g_tm_tI_t-m_t}{m_bI_b}=\frac{m_t}{m_b}\,\frac{g_tI_t-1}{I_b},
	\nonumber \\
	g_b &=& \frac{m_b+g_{tb}m_tI_t}{m_bI_b}=\frac{1}{I_b}\left(
	1+\frac{m_t^2}{m_b^2}\,\frac{g_tI_t-1}{I_b}I_t \right).
\end{eqnarray}
Now we set $g_t=g/C$, and
using Eq.\ (\ref{Is1}) for the $I$'s with $g\Lambda ^2\gg1$, we find
\begin{eqnarray}
	\label{gnew2}
	g_{tb} &=& \frac{g}{C}\frac{m_t}{m_b}\left( 1-\frac{m_t^2}{\Lambda
	^2}\ln\frac{\Lambda ^2}{m_t^2}+\frac{m_b^2}{\Lambda
	^2}\ln\frac{\Lambda ^2}{m_b^2}+\cdots
	\right)-\frac{m_t}{m_b}\frac{1}{C\Lambda ^2}+{\rm O}\left(
	\frac{1}{\Lambda ^4} \right),\nonumber \\
	g_b &=& \frac{g}{C}\frac{m_t^2}{m_b^2}\left( 1-\frac{2m_t^2}{\Lambda
	^2}\ln\frac{\Lambda ^2}{m_t^2}+2\frac{m_b^2}{\Lambda
	^2}\ln\frac{\Lambda ^2}{m_b^2}+\cdots \right)+\left(
	1-\frac{m_t^2}{m_b^2} \right)\frac{1}{C\Lambda ^2}+\cdots,\nonumber \\
	g_t &=& \frac{g}{C}.
\end{eqnarray}
The determinant of {\bf G} is simple, and is zero up to
O$(1/\Lambda ^2)$,
\begin{equation}
\label{detG3}
	{\rm det}{\bf G}=\frac{g}{C^2\Lambda ^2}\left(
	1+\frac{m_t^2}{m_b^2} \right)+{\rm O}\left( \frac{1}{\Lambda ^4} \right).
\end{equation}
The eigenvalues and eigenvectors of {\bf G} are
\begin{eqnarray}
	 \hspace{1.25in}G_+&=&\frac{g}{C}
	\left( \frac{m_t^2}{m_b^2}+1
	\right)-\frac{2g\ln \Lambda ^2}{C\Lambda
	^2}\frac{m_t^2}{m_b^2}\left( m_t^2-m_b^2
	\right)\nonumber\\
	& &+\frac{m_t^2/m_b^2}{C\Lambda ^2}\left[ -1+2g\left(
	m_t^2\ln m_t^2-m_b^2\ln m_b^2 \right) \right],\nonumber\\
	G_-&=&\frac{1}{C\Lambda ^2},\nonumber\\[3ex]
\label{Aev1}
	\mbox{\boldmath$\xi $}_+&=&\left(
	\begin{array}{c}
		a_+\\
		b_+
	\end{array}
	 \right)=N\left(
	 \begin{array}{c}
	 	1\\
	 	m_t/m_b
	 \end{array}
	  \right)+{\rm O} \left( \frac{1}{C\Lambda ^2} \right) \approx N\left(
	  \begin{array}{c}
	  	1\\ a
	  \end{array}
	   \right),\nonumber\\
	  \mbox{\boldmath$\xi $}_-&=&\left(
	  \begin{array}{c}
	  	a_-\\
	  	b_-
	  \end{array}
	   \right)=N\left(
	   \begin{array}{c}
	   	m_t/m_b\\
	   	-1
	   \end{array}
	    \right)+{\rm O} \left( \frac{1}{C\Lambda ^2} \right)\approx N\left(
	    \begin{array}{c}
	    	a\\ -1
	    \end{array}
	     \right), \nonumber\\
	N &=& \frac{1}{\sqrt{1+m_t^2/m_b^2}}=\frac{1}{\sqrt{1+a^2}}.
	\end{eqnarray}
Thus the eigenvectors for the two cases are the same to leading
order, and the eigenvalue $G_-$ is the same. By inspection, ${\bf
m}\propto\mbox{\boldmath$\xi $}^-$ as required in Eq.\
(\ref{G-,1}); and we see that it is natural for the leading order mass gap
equations, as expressed by Eq.\ (\ref{G-1,2}), to be satisfied for
either solution of the couplings $g_i$.

In this case the mass ratio is (to order $1/\Lambda ^2$),
\begin{equation}
	\label{Amtmb1}
	\frac{m_t}{m_b}=\frac{g_b}{g_{tb}}=\frac{g_{tb}}{g_t}.
\end{equation}
We note that $m_t \gg m_b$ requires $g_b \gg g_{tb} \gg g_t$,
a surprising ordering of the couplings.  This solution is {\em not}
obtainable from the solution of Ref. \cite{bhl}, $g_t=1/C\Lambda ^2,
g_b=g_{tb}=0$.

The dimensionless couplings $g'=g\Lambda ^2$ are, for the old
solution $g'\sim{\rm O}(1)$ and for the new solution $g'\gg1$. The
new solution corresponds to strong coupling, and in the
eigenvector space, the direction $\mbox{\boldmath$xi$}_-$ which goes with the
small eigenvalue $G_-=1/C\Lambda ^2$ is a weakly coupled direction
while the $\mbox{\boldmath$xi$}_+$ direction which goes with the finite-sized
$G_+$ eigenvalue is strongly coupled.
States which are expressed in terms of $\mbox{\boldmath$xi$}_+$ are very
massive, while
states which are expressed
in terms of $\mbox{\boldmath$xi$}_-$ have small masses relative to $\Lambda $.

\section{A new symmetry of the interaction Lagrangian and strong
coupling}

In the new solution for the $g$'s, det{\bf G} is very small, ${\rm
det}{\bf G}={\rm O}(1/\Lambda ^2)+{\rm O}(1/\Lambda ^4)$. We can
investigate what would happen if ${\rm det}{\bf G}=0$, which
corresponds to one eigenvalue being zero. Since $G_-=1/C\Lambda ^2$ and
$G_+=(g/C)(1+m_t^2/m_b^2)$, $G_-$ is the obvious eigenvalue to turn
off. In this regime, ${\cal L}_{\rm int}$ can be written as
$\Psi ^\dagger\Psi $, where $\Psi $ is a mixed neutral and charged
state, though there is nothing particular to be gained by this. We
will label quantities with a zero to designate this regime, {\em e.g.},
${\cal L}_{\rm int}^0$, ${\bf G}_0$, $G_+^0$, $\mbox{\boldmath$xi$}^0_\pm$.

When $G_-$ is turned off, the direction $\mbox{\boldmath$xi$}_-^0$ is ``flat".
States of the Lagrangian proportional to $\mbox{\boldmath$xi$}_-^0$ do not
couple, scatter, or bind in the bubble approximation. Bound states
which still may exist, those proportional to $\mbox{\boldmath$xi$}_+^0$, would
have very large masses $\sim\Lambda $. We think of setting $G_-$,
expressed in terms of $g_t,g_b$, and $g_{tb}$, to zero, {\em
not\/} of taking $\Lambda \rightarrow\infty$. Then when the couplings
are changed slightly so that ${\rm det}{\bf
G}\approx g_tg_b-g^2_{tb}\not=0$, {\em i.e.}, ${\bf G}\not={\bf G}_0$,
the coupling eigenvalue $G_-$ is turned on with value
$1/C\Lambda ^2$, and the quark masses and boson bound states turn on
at a mass scale $\ll\Lambda $. The old solution with all couplings
O$(1/\Lambda ^2)$ used the fine-tuning condition $g\sim{\rm
O}(1/\Lambda ^2)$ to get small enough masses, while the new
solution has a different ``fine-tuning" condition ${\rm det}{\bf
G}\sim{\rm O}(1/\Lambda ^2)$, or ${\rm det}{\bf G}_0=0$,
corresponding to the existence of the flat direction, and
naturally small masses.

We actually discovered the new solution by studying the
expressions for the eigenvalues $G_\pm$ and asking what would
happen if one eigenvalue were zero, $G_-$ being the obvious
choice. That implied ${\rm det}{\bf G}=0$, and we realized that in
principle there was no restriction on the size of $G_+$, hence on
the sizes of all the $g$'s, as long as $G_-$ and ${\rm det}{\bf G}
$ were small, O$(1/\Lambda ^2)$.

The interaction Lagrangian is reduced to
\begin{equation}
	\label{LOint}
	{\cal L}_{\rm int}^0=G_+^0\left[ \left( {\bf
	x}^\dagger\mbox{\boldmath$xi$}_+^0 \right)\left(
\mbox{\boldmath$xi$}_+^{0\dagger}{\bf x}
	\right)+\left( {\bf x}'^\dagger\mbox{\boldmath$xi$}_+^0 \right)\left(
	\mbox{\boldmath$xi$}_+^{0\dagger}{\bf x}' \right) \right].
\end{equation}
The symmetry transformation which leaves ${\cal L}_{\rm int}^0$
invariant is to change any vector by a multiple of $\mbox{\boldmath$xi$}_-$,
\begin{eqnarray}\label{symtx}
	{\bf x} &\rightarrow& {\bf x+\delta x},\nonumber \\
	{\bf \delta x} &=& c(\mbox{\boldmath$xi$}_-^\dagger{\bf
x})\mbox{\boldmath$xi$}_-,\nonumber \\
	{\bf \delta x}^\dagger &=& c\mbox{\boldmath$xi$}^\dagger_-({\bf
	x}^\dagger\mbox{\boldmath$xi$}_-),\nonumber \\
	{\cal L}_{\rm int}^0 &\rightarrow& {\cal L}^0.
\end{eqnarray}
This is not a symmetry of the kinetic part of the Lagrangian, only
of the interaction.

We know that the solution to the mass-gap equations requires ${\rm
det}{\bf G}\not=0$, so $m_t=m_b=0$ initially, and quantities
reduce to simpler forms. Since $g_{tb}^2=g_{t}g_b$, the component
$a$ in $\mbox{\boldmath$xi$}_+^0$, see Eq.\ (\ref{evec1}), is now
\begin{equation}
	\label{a0}
	a^0 = \sqrt{\frac{g_b}{g_t}}.
\end{equation}
The full set of eigenstates $\Phi $ of the Lagrangian, see Eq.\
(\ref{L3}), is
\begin{equation}
\label{estt1}
\begin{array}[b]{l@{\hspace{3em}}l}
	\displaystyle
	\Phi _+^{\rm neu\ s}=\left( \mbox{\boldmath$\xi $}_+^\dagger{\bf
	x} \right)^{\rm s}=\frac{\bar tt+a\bar bb}{(1+a^2)^{1/2}},
	&\displaystyle \Phi _-^{\rm neu\ s}=\left( \mbox{\boldmath$\xi $}_-^
	\dagger{\bf
	x} \right)^{\rm s}=\frac{a\bar tt-\bar bb}{(1+a^2)^{1/2}},\\[3ex]
	\displaystyle
	\Phi _+^{\rm neu\ ps}=\left( \mbox{\boldmath$\xi $}_+^\dagger{\bf
	x} \right)^{\rm ps}=\frac{\bar t\gamma _5t-a\bar b\gamma
	_5b}{(1+a^2)^{1/2}},
	&\displaystyle
	\Phi _-^{\rm neu\ ps}=\left( \mbox{\boldmath$\xi $}_-^\dagger{\bf
	x} \right)^{\rm ps}=\frac{a\bar t\gamma _5t+\bar b\gamma
	_5b}{(1+a^2)^{1/2}},\\[3ex]
	\begin{array}[t]{l}
	\displaystyle
	\Phi _+^{\rm chg}=\mbox{\boldmath$\xi $}_+^\dagger{\bf x}'
	=\displaystyle\frac{2(-a\bar b_Rt_L+\bar b_Lt_R)}{(1+a^2)^{1/2}},
	\end{array}
	&
	\begin{array}[t]{l}
		\displaystyle\Phi _-^{\rm chg}=\mbox{\boldmath$\xi
		$}_-^\dagger{\bf x}'
		=\displaystyle=\frac{2(\bar b_Rt_L+a\bar b_Lt_R)}{(1+a^2)^
	{1/2}}.
	\end{array}
\end{array}
\end{equation}
The $\Phi _-$ states are the ones corresponding to composite bound
state Higgs ($\Phi _-^{\rm neu\ s}$), neutral Goldstone ($\Phi
_-^{\rm neu\ ps}$), and charged Goldstone ($\Phi _-^{\rm chg}$)
bosons. One way to view the breaking of the new symmetry is that
as $G_-$ turns on from 0 to $1/C\Lambda ^2$, the $a$ in these states
changes slightly from $-g_{tb}/g_t=-\sqrt{g_b/g_t}$ to
$-g_{tb}/(g_t-G_-)$.

We do not yet understand this new symmetry in any fundamental way.

\section{Summary of results}

Every two-quark result presented in this paper generalizes
straightforwardly to $2n$ quarks. For example, there would be
$(2n-1)$ large eigenvalues and one small one for ${\bf G}$, ${\rm
det}({\bf GI-1})=0$, and ${\rm det}({\bf GJ-1})=0$. We have checked
the charged Goldstone boson bound-state condition and find the
same type of {\em ad hoc} rules to impose for $2n$ quarks as for two quarks.
In the
$G_-=0$ regime, there would be $(2n-1)$ times as many high-mass
bound states, but the same ``flat" states.

Our most important new result is the new solution for the
couplings $g_i$ which allow the mass-gap equations to be
satisfied, presented in Section III. These couplings correspond to
dimensionless couplings $g_i'=g_i\Lambda ^2\gg1$, instead of
$g_i'\sim{\rm O}(1)$ as in the previous solutions with fine-tuning
condition $g_i\sim{\rm O}(1/\Lambda ^2)$. Despite the strong
coupling, the existence of a nearly flat direction in the interaction
gives naturally small masses $m_t,m_b,m_H\ll\Lambda $.

The new solution grew out of the matrix formulation of Section II.
In that section, we pointed out some details not previously
discussed in the literature. Even for the old solution with
$g_i\sim{\rm O}(1/\Lambda ^2)$, the next-to-leading terms are
important for getting exact solutions to the mass-gap equations
and bound-state conditions. Since $G_-=1/C\Lambda ^2$ in both
solutions for the $g_i$, our determinant arguments hold for both
solutions. The non-vanishing mass in the charged Goldstone boson
case is a consequence of the one-loop approximation and can be
fixed by setting $m_t$ or $m_b=0$ in loops, depending on which
coupling and which connecting states are involved.

The new solution led to the realization of a new symmetry of
${\cal L}_{\rm int}$, discussed in Section V. Breaking this
symmetry appears to have the same result as breaking $\rm
SU(2)_L\times U(1)_R$, yet it is not a symmetry of the whole
Lagrangian. When the eigenvalue $G_-$ is zero, the relevant bound
states are in the flat $\mbox{\boldmath$xi$}_-$ direction. When $G_-$ is turned
on with value $1/C\Lambda ^2$, these states appear as the Higgs
and three Goldstone bosons, and the quarks acquire masses.

We believe that our solution should be of interest in the mass
condition $m_t\ll\Lambda \sim2$ TeV, which corresponds to one
model (technicolor).

\acknowledgements

Most of this work was done while BD was a participant at the Aspen
Center for Physics. Loyal Durand provided invaluable comments as
the work progressed.  This work was supported in part by the U.S. Department
of Energy under Contract No. AC02-76ER0081.

\end{document}